\documentclass{pos}
\usepackage{epsfig,graphicx}
\usepackage{amssymb}
\usepackage{cite}

\newcommand{\beq}{\begin{equation}}
\newcommand{\eeq}{\end{equation}}
\newcommand{\be}{\begin{equation}}
\newcommand{\ee}{\end{equation}}
\newcommand{\ba}{\begin{array}}
\newcommand{\ea}{\end{array}}
\newcommand{\beqa}{\begin{eqnarray}}
\newcommand{\eeqa}{\end{eqnarray}}
\newcommand{\bea}{\begin{eqnarray}}
\newcommand{\eea}{\end{eqnarray}}
\newcommand{\lsim}{\stackrel{<}{_\sim}}
\newcommand{\gsim}{\stackrel{>}{_\sim}}
\newcommand{\cA}{{\cal A}}
\newcommand{\cO}{{\cal O}}

\newcommand{\cL}{{\cal L}}
\newcommand{\Tr}{{\rm Tr}}

\newcommand{\no}{\nonumber}

\title{Effective theories of electroweak symmetry breaking}

\ShortTitle{EFT of electroweak symmetry breaking}

\author{\speaker{Gino Isidori}\thanks{Work supported by the EU under contract 
MTRN-CT-2006-035482.}\\
        INFN, Laboratori Nazionali di Frascati, Via E. Fermi 40
I-00044 Frascati, Italy\\
 Institute for Advanced Study, 
 T.~U.~M\"unchen, Arcisstra\ss e 21, D-80333 M\"unchen, Germany\\        
E-mail: \email{Gino.Isidori@lnf.infn.it}}

\abstract{We present a brief review of recent 
attempts to construct effective theories to describe the breaking 
of the electroweak symmetry in extensions of the
Standard Model with new strongly interacting dynamics 
around the TeV scale. Particular attention is devoted 
to the analogies of  
Higgsless models with Chiral Perturbation Theory
(the low-energy limit of QCD).}

\FullConference{6th International Workshop on Chiral Dynamics, CD09\\
		July 6-10, 2009\\
		Bern, Switzerland}

\begin{document}

\section{Introduction}

Despite the impressive phenomenological success of the Standard Model (SM), 
our understanding of one of its key ingredients, namely the 
spontaneous breaking of the $SU(2)_L\times U(1)_Y$ gauge symmetry, 
is still very poor. We have very solid evidences that 
$SU(2)_L\times U(1)_Y$ is a local symmetry of the underlying theory, 
and that this symmetry is only spontaneously broken. However, the fact
that this breaking is induced by the non-trivial vacuum expectation 
value of a single  $SU(2)_L$ scalar field (the Higgs field), 
is far from being clearly established.  

A fundamental Higgs boson is certainly the most economical
way to explain the $SU(2)_L\times U(1)_Y \to U(1)_Q$ breaking, 
and a light Higgs mass ($m_{h}\approx $~100~GeV) is also an efficient 
way to account for all the existing electroweak precision tests. 
However, on the one hand we have no direct evidences of the physical 
Higgs boson. On the other hand, the strong sensitivity of $m_{h}$ to 
short-distance scales poses a serious naturalness problem to the theory. 

Several alternative symmetry-breaking mechanisms have been 
proposed in the literature, from Higgsless theories, to models 
with composite or partially-composite Higgs bosons. As we will 
discuss in the following, the low-energy features of all these
models are universal and can be described in general terms 
using appropriate effective theory approaches.

\section{A closer look to the {\em Standard} Higgs sector}

Before starting the discussion of alternative scenarios, 
it is convenient to give a closer look to the 
standard Higgs mechanism. The symmetry-breaking 
part of the SM Lagrangian can be written as follows
\beqa
\cL^{\rm SM}_{\rm S.B.} &=& \cL_{\rm Higgs} + \cL_{\rm Yukawa}~, \\
\cL_{\rm Higgs} &=& 
\frac{1}{4} \Tr \left( D_\mu H^\dagger D_\mu H \right) 
+ \frac{\mu^2}{4} \Tr \left(H^\dagger H \right) 
- \frac{\lambda}{16} \left[\Tr \left(H^\dagger H \right) \right]^2~.
\label{eq:LHiggs}
\eeqa
Here we have expressed the Higgs field in terms of the 
$2\times 2$ matrix $H$, defined by $H = ( i \sigma_2 \phi^*, \phi)$,
in terms of the familiar $SU(2)_L$ complex doublet $\phi$:
\bea
&& \phi = \left(\ba{c} \phi_1 +i \phi_2  \\ \phi_3 +i \phi_4 \ea \right)~, \qquad 
H = ( i \sigma_2 \phi^*, \phi)~, \\
&& D_{\mu}H = \partial_{\mu}H  - i  g T_{a}W_{\mu}^{a}H + i  g^{\prime} H T_{3}B_{\mu}~,
\label{eq:Dmu}
\qquad 
T_{a}=\frac{1}{2}\sigma^{a}~.
\eea
This notation has the advantage of making manifest the invariance of the 
SM Higgs potential under the {\em global}  $SU(2)_L \times SU(2)_R$ symmetry, 
defined by $H \to g_L H g^\dagger_R$. This global symmetry is spontaneously broken 
into the diagonal subgroup $SU(2)_{L+R}$ (often known as {\em custodial} symmetry)
by the Higgs vacuum expectation value: 
$\langle H \rangle = v \times I$, which implies the tree-level relations 
\be
m^2_W = \frac{g^2 v^2}{4}~, \qquad m^2_Z = \frac{m_W^2}{\cos^2\theta_W}~, 
\label{eq:mWZ}
\ee
for $W$ and $Z$ boson masses  ($\cos^2\theta_W = g^2/(g^2+g^{\prime2})$,
$\left. v^2 \right|_{\rm tree} = \mu^2/\lambda)$.

As shown 
by the expression of the covariant derivative in (\ref{eq:Dmu}),
the $SU(2)_L$ subgroup of this global symmetry is fully gauged, 
while only the $T_3$ component of  $SU(2)_R$ is gauged, giving rise to 
an explicit breaking of the custodial symmetry beyond the tree-level.
More precisely, $T^R_3$ allows to write  the generator of the $U(1)_Y$ 
gauge group as
\be
Y = T^R_3 + \frac{1}{2} (B-L)
\ee
where $B$ and $L$ are the barion- and lepton-number generators
(which appear in the covariant derivatives acting on 
the fermion fields). An additional source of breaking of the custodial symmetry is present 
in the Yukawa interaction, due to the difference between up- and down-quark
Yukawa couplings:
\be
\cL^{\rm quark}_{\rm Yukawa} ~=~   \frac{1}{2}(Y_U^{i} + Y_D^{i})\bar Q^i_L H Q^j_R  ~+~
(Y_U^{i} - Y_D^{i}) \bar Q^i_L  H T_3 Q^j_R  + {\rm h.c.},
\qquad Q^i_{L(R)} = \left( \ba{c} u_{L(R)}^i \\ d_{L(R)}^i \ea \right)~.
\ee

All the physical parameters appearing in $\cL^{\rm SM}_{\rm S.B.}$ have
been determined with high accuracy but for the combination controlling 
the mass of the physical Higgs boson:  $m_h = 2 \mu^2 = 2 \lambda v^2$,
at the tree level. In other words, we have a good determination of the 
ground state determined by $\cL^{\rm SM}_{\rm S.B.}$ but only 
a limited knowledge of the dynamics behind the symmetry breaking mechanism.

\begin{figure}
\begin{center}
\includegraphics[width=200pt]{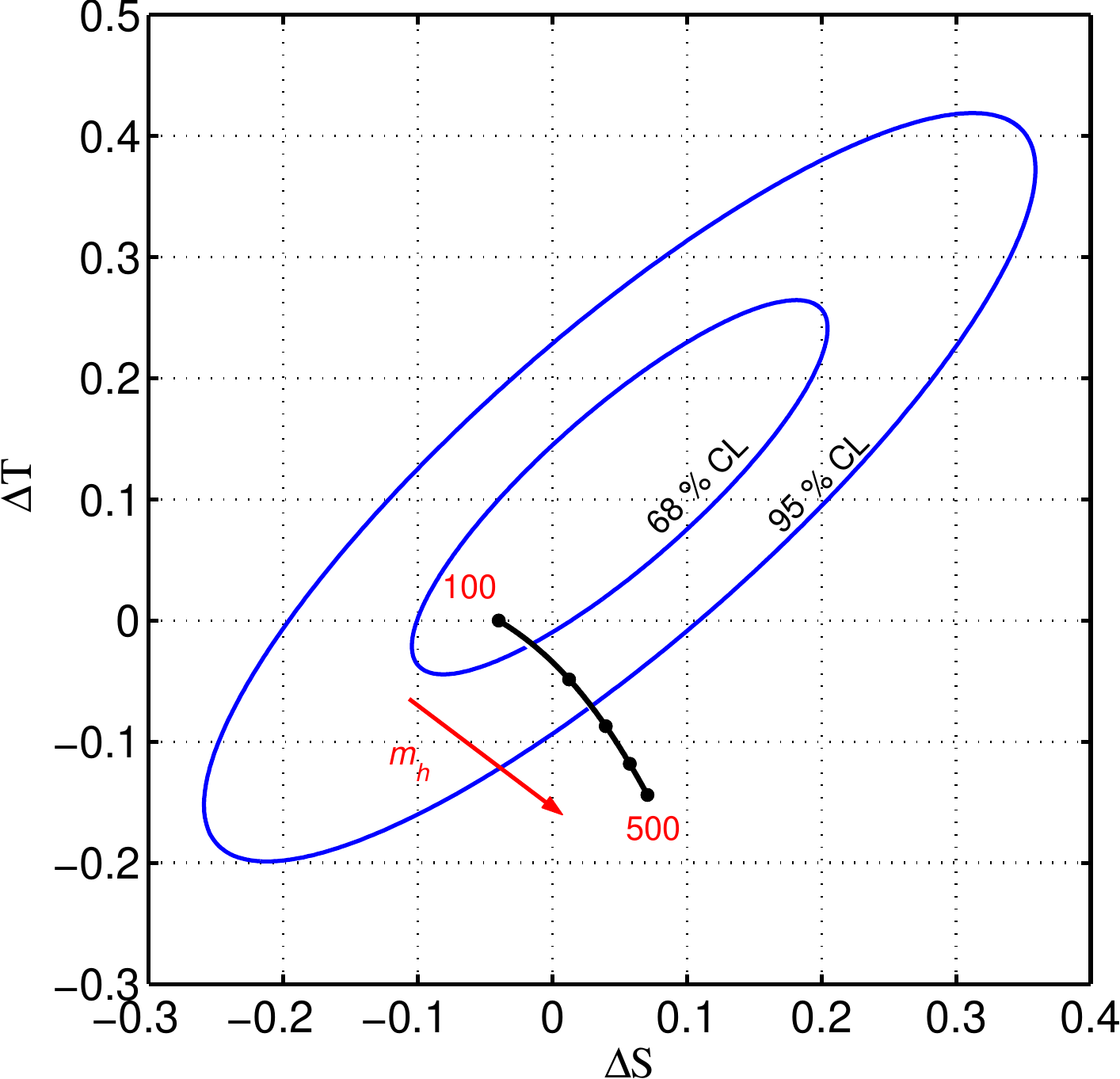}
\end{center}
\caption{Experimentally allowed range for the $S$ and $T$ parameters
(blue ellipses), from Ref.~\cite{Barbieri:2007gi}. 
The $\Delta S=\Delta T=0$ point corresponds to the SM 
prediction for $m_t=175$~GeV and $m_h=100$~GeV. The black curve is
the SM prediction for  $m_t = 171.4$~GeV and different values of 
$m_h$ (in GeV). }
\label{fig:ST}
\end{figure}

A non-trivial sensitivity to the Higgs mechanism is obtained 
from the Electroweak Precision Observables (EPWO), and in particular  
by the two-point effective couplings controlling masses and 
wave-functions of the massive gauge-bosons,
\be
S = \frac{16 \pi}{g g^\prime} \left. \frac{\partial}{\partial q^2} 
\cA(W_3 \to B)\right|_{q^2=0}
\qquad {\rm and} \qquad
T = \frac{1}{\alpha} \left( \frac{m^2_W }{m_Z^2 \cos^2\theta_W}  -1 \right)~.
\ee
Within the SM these two effective couplings can be 
predicted unambiguously in terms of the free parameters of 
$\cL^{\rm SM}_{\rm S.B.}$ (including $m_h$),
while experimental data allows us to determine them 
independently of any assumption on $m_h$. The comparison 
of data vs.~theory (including all the relevant quantum corrections) 
is shown in Fig.~\ref{fig:ST}. From this plot it is clear that 
a light Higgs boson is an economical and consistent way to 
accommodate existing data. However, it must be stressed 
that this conclusion holds under the hypothesis of an 
heavy cut-off for the SM viewed as an effective theory. 

If we extend the theory to include new degrees degrees 
of freedom above a cut-off scale $\Lambda > v$ ($v\approx 246$ GeV), we can view 
$\cL^{\rm SM}$ as the low-energy limit of an effective Lagrangian, 
with an infinite tower of higher-dimensional operators suppressed
by inverse powers of  $\Lambda$. In this perspective, 
we should expect corrections to $S$ and $T$ by 
operators of the from
\be
Q_S = g g^\prime \Tr\left( H T_3 B_{\mu\nu} H^\dagger T_{a}W_{\mu\nu}^{a} \right)~,
\qquad Q_T = e^2 \left[\Tr\left( T_3 H^\dagger D_\mu H \right)
\right]^2~,
\label{eq:QSQT}
\ee
appearing in the dimension-six part of the Lagrangian,
$\cL_{d=6} = (c_T/\Lambda^2) Q_T + (c_S/\Lambda^2) Q_S + \ldots$, 
namely
\be
\Delta S = - c_S \frac{ 16 \pi v^2}{\Lambda^2}~, \qquad 
\Delta T =   c_T \frac{ 8 \pi v^2}{\Lambda^2}~.
\label{eq:STloc}
\ee
Assuming $c_{S,T}=\cO(1)$, as expected by na\"ive dimensional analysis, 
we conclude that the indication of a light Higgs ($m_h \lsim 200$ GeV) 
holds for $\Lambda \gsim 4$~GeV. 

Since $m_h$ (or the dimension-2 operator in $\cL_{\rm Higgs}$) 
is quadratically sensitive to the cut-off,
the hierarchy between $\Lambda$ and $m_h$ poses a naturalness 
problem to the theory (the so-called {\em little hierarchy} problem).

\section{Breaking the electroweak symmetry without the Higgs}

If we assume that the electroweak symmetry is  broken
because of an underlying spontaneous breaking of a 
global symmetry, the low-energy dynamics of the system is universally  
determined by the properties of the corresponding 
Goldstone bosons~\cite{Coleman:1969sm}.  If we assume that the underlying 
symmetry is $SU(2)_L \times SU(2)_R \to SU(2)_{L+R}$,
as in the standard case, the bosonic part of the low-energy 
effective Lagrangian has a unique operator:
\be
\cL_\chi^{(2)} =  \frac{v^2}{4} \Tr \left( D_\mu U^\dagger D_\mu U \right)~,
\ee
where $U$ is the unitary matrix encoding 
the three Goldstone bosons fields ($\pi_i$):
\be
U=e^{2i\hat{\pi}/v},\qquad\hat{\pi}=T^{a}\pi^{a}=\frac{1}{\sqrt{2}}\left[
\begin{array}
[c]{cc}%
\frac{\pi^{0}}{\sqrt{2}} & \pi^{+}\nonumber\\
\pi^{-} & -\frac{\pi^{0}}{\sqrt{2}}%
\end{array}
\right]~, \qquad U\to g_L U g_R~.
\ee
Not surprisingly, we obtain $\cL_\chi^{(2)}$ also from
$\cL_{\rm Higgs}$ in the limit $m_h \to \infty$, as it 
can easily be check from (\ref{eq:LHiggs}) re-writing $H$ as
\be
H = (v + h) \times U~.
\ee
In principle, to break the electroweak symmetry we do not need 
to assume an underlying $SU(2)_L \times SU(2)_R$ global symmetry: 
a $SU(2)_L \times U(1)_R$ group would be sufficient.  
However, doing so the universal $\cO(p^2)$ Lagrangian
of the Goldstone bosons would contain also the
custodial-symmetry-breaking operator 
\be
O^{(2)}_{T}  ~=~  \frac{v^2}{4} \left[ \Tr \left( T_3 U^\dagger D_\mu U \right) \right]^2~.
\ee
Since this operator breaks the relation between $W$ and $Z$ 
masses in (\ref{eq:mWZ}) it must have an unnaturally small coefficient. 
This is why a global $SU(2)_L \times SU(2)_R$ symmetry 
(or any larger group containing $SU(2)_L \times SU(2)_R$) 
is the natural starting point for an effective description 
of the electroweak symmetry breaking mechanism.

The effective Lagrangian 
\be
\cL^{\rm Univ}_{\rm eff}  =  \cL^{\rm SM}_{\rm gauge} + \cL_\chi^{(2)} 
+ \cL_{\rm Yukawa} ( H \to v\times U)~.
\label{eq:Luniv}
\ee
provides an excellent description of particle physics,
beyond the tree level,
at energies below the ultraviolet cut-off 
\be
\Lambda_\chi = 4 \pi v \approx 3~{\rm TeV}
\ee
(the na\"ive dimensional cut-off of $\cL_\chi^{(2)}$).
Most low-energy precision observables (including the anomalous 
magnetic moment of the muon, $B\to X_s\gamma$, $\Delta M_B$, etc\ldots)
can be computed to high precision with this effective Lagrangian
and are in agreement with experimental data.
There are only two potential problems, which are usually advocated 
as clues of new degrees of freedom below the cut-off $\Lambda_\chi$:
\begin{itemize}
 \item the violation of unitary in $WW$ scattering, if evaluated at the 
 tree-level with $\cL^{\rm Univ}_{\rm eff}$;
 \item the bad agreement with data of the electroweak observables $S$ and $T$,
 if evaluated at the one-loop level with $\cL^{\rm Univ}_{\rm eff}$, using 
 $\Lambda_\chi$ as ultraviolet cut-off.
\end{itemize}

The tree-level violation of unitary in $WW$ scattering cannot be considered 
a clear indication of new degrees of freedom below the scale  $\Lambda_\chi$. 
We find exactly the same problem in $\pi\pi$ scattering within QCD.
As in the QCD case, 
the problem could be solved by strongly interacting dynamics 
around the na\"ive ultraviolet cut-off of the effective theory. 
Only if the ultraviolet completion 
of the theory is weakly interacting, we are forced to include {\em light}
degrees of freedom below $\Lambda_\chi$.
This is what happens with the standard Higgs mechanism, 
where the new degrees of freedom is the physical Higgs field.

The problem of $S$ and $T$ is the most difficult one to be solved. 
$\cL^{\rm Univ}_{\rm eff}$ allows us to evaluate the universal
infrared contribution to $S$ and $T$ generated by Goldstone-boson
loops. Contrary to the flavour observables mention above, 
this contribution is not finite. At the one-loop level 
one has 
\be
\left.  \Delta T\right\vert_{\rm IR} =-\frac{3}{8\pi c_{W}^{2}}
\ln\left(  \frac{\Lambda}{m_{W}}\right)+\ldots~, \qquad 
\left.  \Delta S\right\vert_{\rm IR} =\frac{1}{6\pi}
\ln\left(  \frac{\Lambda}{m_{W}}\right)+\ldots~
\label{eq:STIR}
\ee
where $\Lambda$ is an ultraviolet regulator and  
the dots are tiny finite corrections. 
Replacing $\Lambda$ with  $\Lambda_\chi$ leads to 
a bad description of experimental data. This can 
be understood by noting that if we replace 
$\Lambda$ with $m_h$ we obtain, to a good accuracy, 
the SM contribution shown in Fig.~\ref{fig:ST} as a function of $m_h$.
Additional contributions to $S$ and $T$ are generated by 
local operators of higher order in the derivative 
expansion. However, such contributions are naturally 
subleading with respect to the logarhythmically-enhanced 
infrared terms in Eq.~(\ref{eq:STIR}): the na\"ive size of the 
local terms is what we obtain  
from (\ref{eq:STloc}) replacing  $\Lambda$  with  $\Lambda_\chi$.
In order to be consistent with data we need a mechanism to 
enhance the local contributions: this can easily be obtained 
including new states below the cut-off scale $\Lambda_\chi$.

\section{{\em Light} vectors in the electroweak 
chiral Lagrangian }

An interesting alternative to the Higgs mechanism is the wide 
class of {\em Higgsless} models: theories where there are no 
light scalar particles and the $SU(2)_L \times U(1)_Y \to U(1)_Q$ 
breaking is generated by new strong dynamics 
above the electroweak scale 
(see e.g.~Ref.~\cite{Casalbuoni:1985kq,Csaki:2003dt,Nomura:2003du,
Barbieri:2003pr,Foadi:2003xa,Chivukula:2004pk,Georgi:2004iy,Foadi:2007ue}).
A general feature of these models is that the lightest 
non-standard particles are massive spin-one states.
These new vectors replace, partly or completely, 
the Higgs boson in maintaining unitarity in $W W \to W W$ 
scattering~\cite{Chivukula:2003kq,Bagger:1993zf},
and may play a significant role also in EWPO.

The phenomenology of new massive vectors at high-energy 
colliders~\cite{
He:2007ge,Accomando:2008jh,Belyaev:2008yj},
as well as their role in electroweak observables, has been widely discussed 
in the recent literature. However, most of the existing analyses 
are based on specific dynamical assumptions, such as considering 
these vector states as the massive gauge bosons of a hidden 
local symmetry. As recently discussed in~\cite{Barbieri:2008cc}
(see also~\cite{Hirn:2007we}), 
these assumptions may be too restrictive for generic models with 
strong dynamics at the TeV scale, and only going beyond these 
assumptions the sole exchange of heavy vectors can provide 
a successful description of EWPO. 
More generally, the construction of an appropriate 
effective theory including only SM fields and these new 
light states is a very efficient tool to discuss theoretical 
and phenomenological constraints on such states. 

A simple prototype is the effective theory proposed 
in~\cite{Barbieri:2008cc}, which is based only on 
the following rather general assumptions:
\begin{itemize}
\item The new strong dynamics is invariant under 
a global chiral symmetry, $SU(2)_L \times SU(2)_R$, broken 
spontaneously into $SU(2)_{L+R}$,
and under the discrete parity $P:\ SU(2)_L \leftrightarrow SU(2)_R$. 
\item A pair of vector ($V$) and axial-vector ($A$) states,
belonging to the adjoint representation of $H$, are the only new {\em light} 
dynamical degrees of freedom below $\Lambda_\chi$. 
\item SM fermions couple to the new $V$ and $A$ states only 
via the SM gauge interactions.
\end{itemize}
Under these assumptions, the low-energy interactions of the 
new vector fields and the SM degrees are controlled by five 
parameters: two masses ($M_V,A$) and three effective couplings. 
The latter are the couplings of the three $\cO(p^2)$ operators 
containing at most one heavy field, 
\be
\mathcal{L}_{1V}^{(2)} = \frac{i}{2\sqrt{2}}G_{V}\Tr\left(
  V^{\mu\nu}[u_{\mu},u_{\nu}] \right) 
 + \frac{1}{2\sqrt{2}}F_{V}\Tr\left( V^{\mu\nu} f^+_{\mu\nu} \right)
 + \frac{1}{2\sqrt{2}}F_{A}\Tr\left( A^{\mu\nu} f^-_{\mu\nu} \right)~,
\label{eq:LV1}
\ee
where 
\be
f^{\pm}_{\mu\nu} = u (g T^a W^a_{\mu\nu})u^{\dagger} \pm 
u^{\dagger} (g^\prime T_3 B_{\mu\nu}) u~,  \qquad 
u_{\mu}=iu^{\dagger}D_{\mu}Uu^{\dagger}~, \qquad u^2 =U~,
\ee
Here we have chosen to describe the new massive vectors 
via the irreducible antisymmetric tensor fields, 
$V^{\mu\nu}$ and $A^{\mu\nu}$, following the 
formalism proposed in Ref.~\cite{Ecker:1989yg}. This formulation 
has the advantage of avoiding the kinetic mixing 
of the new states with the Goldstone boson fields, and
allows the most general form of interaction at $\cO(p^2)$.
Indeed this Lagrangian is nothing but the translation 
to high energies of the formalism used in~\cite{Ecker:1988te} 
to describe the low-energy effects of vector-meson 
dominance in QCD.

The three effective couplings, $F_{V,A}$ and $G_V$, have dimensions of mass
and, by naive dimensional analysis, are expected to be of $\cO(v)$.
In specific Higgsless models, such as the so-called 3- and 4-site 
models, these couplings are not independent and obey the 
``hidden-gauge'' relations $F_{V}=2G_V$ and $F_A=0$. These relations 
holds in all five-dimensional deconstructed models as long as 
we neglect non-renormalizable terms in the bulk; however, 
as discussed in~\cite{Hirn:2007we}, they can be violated in 
more general Higgsless frameworks. The effective Lagrangian 
in Eq.~(\ref{eq:LV1}) has the advantage of allowing 
the description of both hidden-gauge models and more 
general frameworks.

\begin{figure}[t]
\begin{center}
\includegraphics[width=7.4cm]{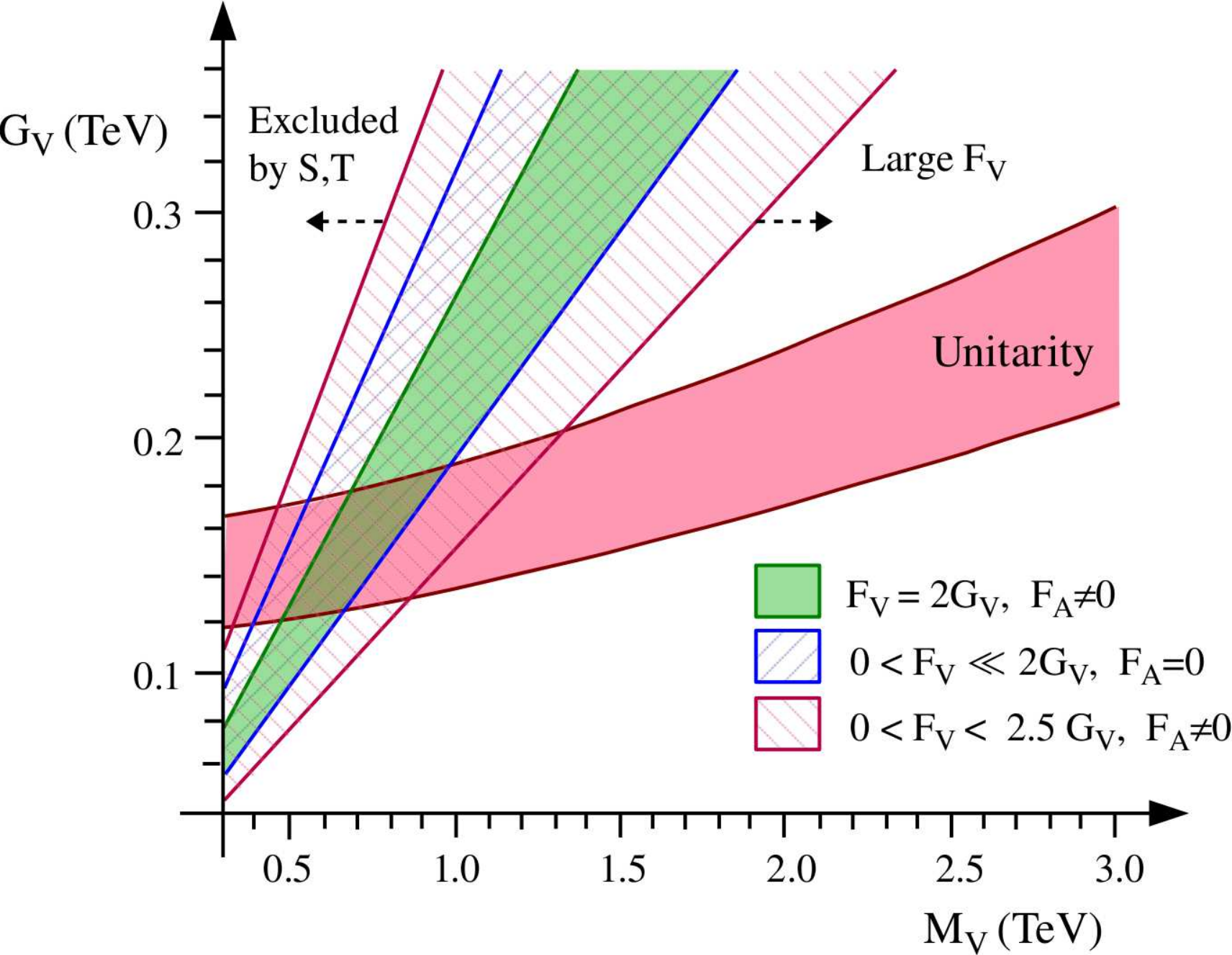}
\includegraphics[width=7.4cm]{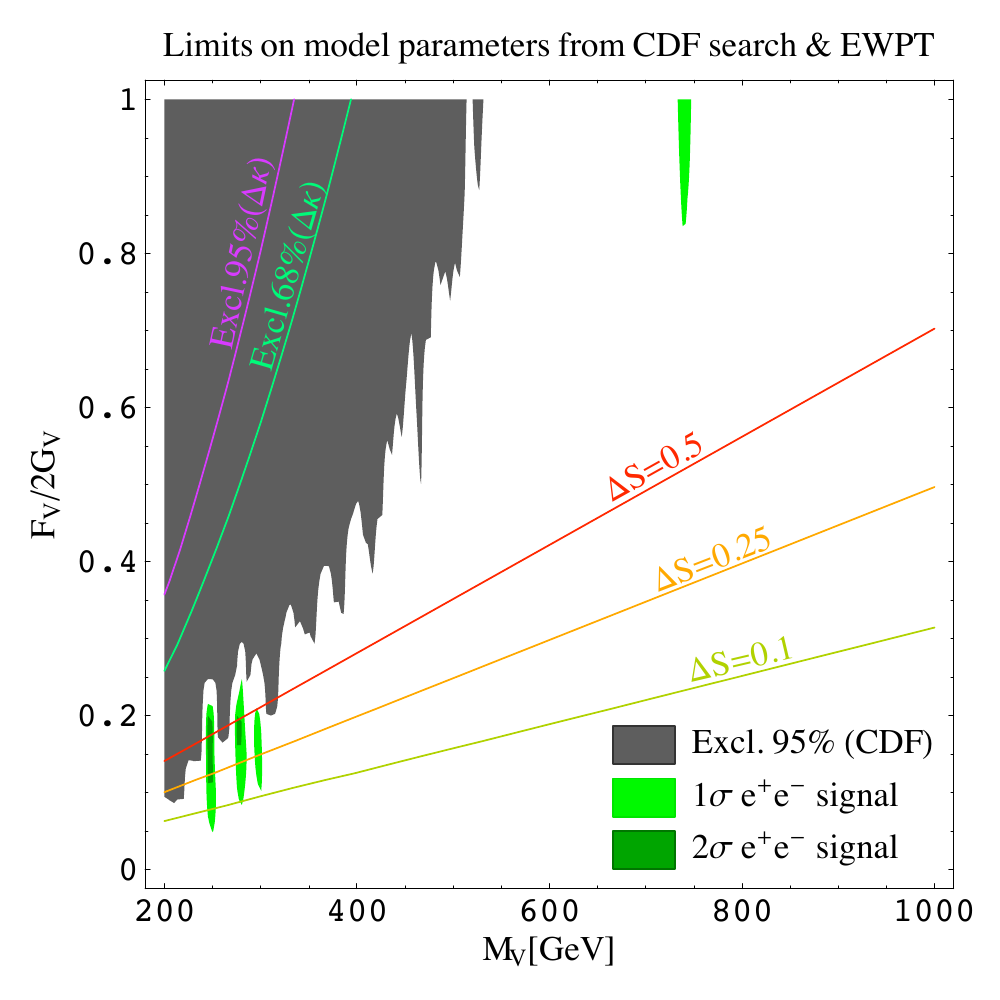}
\end{center}
\caption{Left: summary of unitarity and EWPO constraints (at 95\% C.L.) in the
$(M_{V},G_{V})$ plane~\cite{Barbieri:2008cc}. 
Right: bounds in the $F_V$--$M_V$ plane from $e^+e^-$ CDF data,
and comparison with some of the constraints from EWPO~\cite{Cata:2009iy}.
\label{fig:GVMV} }
\end{figure}

The coupling $G_V$, which controls the coupling of the vector
state to two longitudinal SM gauge bosons, can be rather 
constrained if we impose perturbative unitarity up to $\Lambda_\chi$. 
Imposing this condition, $G_V$ has to lie in the narrow range 
shown in Fig.~\ref{fig:GVMV} (left). Note that this condition does 
do not pose a significant constraint on the value of $M_V$:
as already discussed in the previous section, the unitarity 
condition do not necessarily imply the existence of 
new light states. 
 
A more constrained picture is obtained 
if we require that the sole exchange of the two spin-one states 
leads to a satisfactory description of EWPO. This can be 
achieved, with a proper tuning of the free parameters, 
if at least one vector state is relatively light~\cite{Barbieri:2008cc}.
As is well known, at the tree-level the $S$ parametr receives a positive 
local contribution
\be 
\Delta S = 4\pi \left(\frac{F_V^2}{M_V^2} -\frac{F_A^2}{M_A^2}\right)~,
\ee
which is the analog of the vector-meson contribution to the 
low-energy coupling $L_{10}$ in QCD~\cite{Ecker:1988te}.
This effect alone would worsen the agreement with data;
however, it can possibly be compensated by a  
sizable (quadratically divergent) positive contribution 
to $T$ generated at one-loop level by the $U(1)_Y$ gauge
interaction. This contribution is sizable,
and the EWPO can be satisfied, only if 
$M_V$ is sufficiently light [see Fig.~\ref{fig:GVMV} (left)] 
and, at the same time, the effective couplings in Eq.~(\ref{eq:LV1})
do not satisfy the hidden-gauge relations.

It is worth to stress that a violation of the 
hidden-gauge relations with a light $M_V$ 
is a  non-trivial price to pay in terms of
naturalness of the effective theory.
The bad ultraviolet behavior implied by the 
violation of the hidden gauge relations makes quite 
unnatural to keep $M_V$ relatively well below the 
$4\pi v$ cut-off. For comparison, we report here
the effective low-energy couplings describing 
the dynamics of vector mesons in QCD, rescaled 
to the corresponding vacuum expectaation value 
$(v\approx 246~{\rm GeV} \to 93~{\rm MeV} \equiv F_\pi)$ : 
\bea
&
F^{(\rho)}_V  = (1.66 \pm 0.01) \times v~,
\qquad 
G^{(\rho)}_V = (0.65 \pm 0.08)\times v~, 
\qquad 
F^{(a_1)}_A   = (1.29 \pm 0.27)\times v~, &
\no \\
&
M^{(\rho)}_V =  0.66 \times (4\pi v)~, 
\qquad
M^{(a_1)}_A = 1.05 \times (4\pi v)~. 
&
\eea
As can be seen, in this case we have sizable devitions
from the hidden gauge relations; however, the vector 
masses are close to the na\"ive cut-off.

\medskip

Determining the free parameters of this effective theory 
from data, via the direct production of the new states at 
colliders is a key ingredient to test the validity of this 
construction. The most general signature of these states
is their appearance in $WW$ (or $WZ$) scattering. 
This effect is related to the role played by 
the new states in unitarizing the theory and,
to a large extent, it can be predicted in a 
model-independent way~\cite{Bagger:1993zf}.
The only relevant free parameter is the mass 
of the lightest vector state, which is not fixed 
by the unitarity condition. As shown by recent 
analyses (see e.g.~\cite{He:2007ge}), 
detecting such states in $WW$
scattering at the LHC is not an easy task:
even for $M_V\sim 700$~GeV an integrated 
statistics of $\cO(100~{\rm fb}^{-1})$ is needed.

If the mass of at least one of the new spin-one 
resonances is relatively low (around or below 1~TeV) 
then such states could be detected 
more easily via Drell-Yan production:
despite the faster drop of the signal/background
ratio for rising $M_{V(A)}$, compared to $WW$ fusion, 
in a large fraction of the parameter space 
the Drell-Yan production may yield a rather large 
and clean non-standard signal, even for 
integrated statistics of $\cO(1~{\rm fb}^{-1})$~\cite{Cata:2009iy}.
In addition to the mass spectrum, the key parameters 
here are the effective couplings $F_{V(A)}$, which 
parametrise the (gauge-invariant) mixing of the new states 
with the SM gauge bosons. Interestingly, the determination 
of these parameters could shed more light on the role 
of the resonances in the EWPO.

\begin{figure}[t]
\begin{center}
\includegraphics[width=9cm]{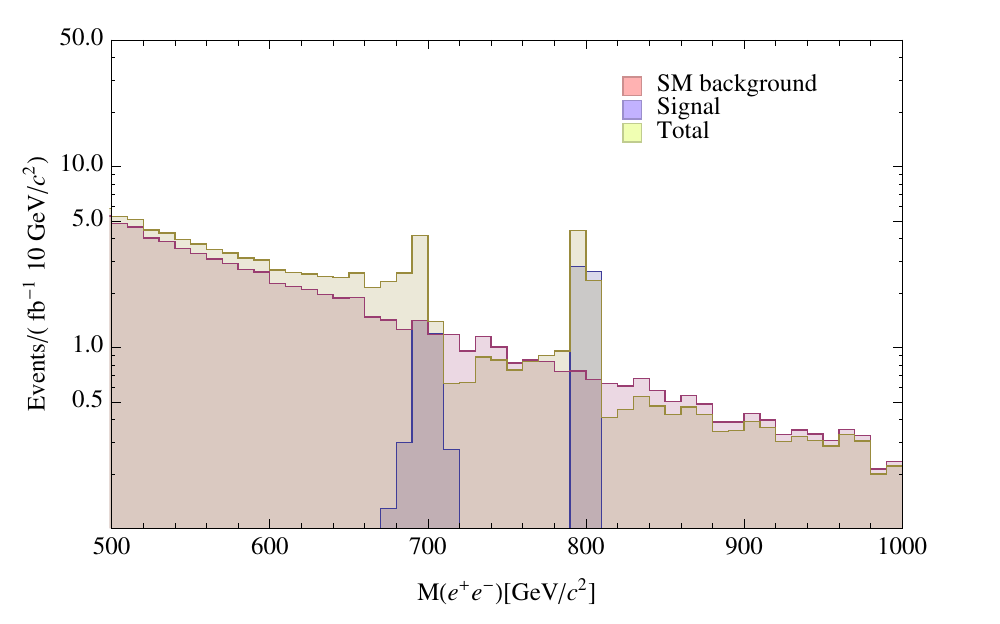}
\includegraphics[width=9cm]{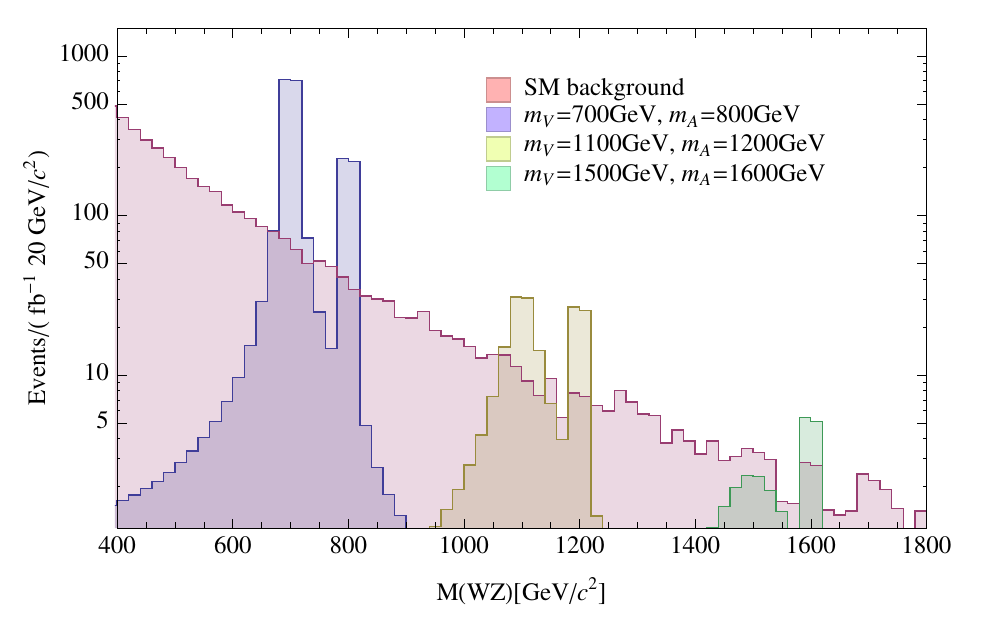}
\vskip -0.5 true cm
\end{center}
\caption{Possible signatures of $V$ and $A$ 
states in $pp \to \ell^+\ell^-$ (up) 
or $pp \to WZ$ (down) at $\sqrt{s}=14$~TeV~\cite{Cata:2009iy}.
All resonance signals have been obtained 
assuming $F_A=F_V=2G_V$ (condition that maximise the 
signal). The SM background corresponds 
only to the irreducible electroweak production of  $\ell^+\ell^-$
or $WZ$ pairs. The plots do not include the decay branching ratios 
of $W$ and $Z$ bosons (lower plot), as well as any 
experimental cut.
\label{fig:WZ} }
\end{figure}

For very light masses ($M_V \lsim 800$~GeV), 
the cleanest signal is the $\ell^+\ell^-$ final state
($\ell=e,\mu$).  
In this channel Tevatron is already providing significant 
constraints in the $F_V$--$M_V$ plane, as shown in 
Fig.~\ref{fig:GVMV} (right).
Note  that relatively low $M_V$ values are still 
allowed, provided $F_V$ is not maximal: a configuration 
which is not allowed in the simplest Higgsless models, 
but is possible (and even favoured by the EWPO) 
in a  general effective theory approach.  
For large values of $F_V$  ($F_V\approx 2G_V$),
there are realistic chances to observe
deviations from the SM at the LHC, even with a statistics 
of a few ${\rm fb}^{-1}$ [Fig.~\ref{fig:WZ} (up)].
However, in this channel 
the signal/background ratio drops very fast with $M_{V}$:
this implies that it is almost impossible 
to detect a signal for $M_V \gsim 800$~GeV
(even with high statistics).
The $WZ$ final state could offer a wider mass reach, 
for sufficiently high statistics. 
As shown in Fig.~\ref{fig:WZ} (down), 
the ratio between signal and irreducible background 
is large even for $M_V \sim 1.2$~TeV. Requiring 
leptonic decays of both $Z$ and $W$, to suppress the 
non-irreducible background, the mass region 
$M_V \gsim 1$~TeV could be explored with an
integrated statistics of $\cO(100~{\rm fb}^{-1})$.

As recently dscussed in Ref.~\cite{Carcamo},
another channel which could be accessile for 
light masses and could be useful to distinguish 
the different realisations of this framework
is the production at the LHC of two massive vectors.

\section{The strongly-interacting light Higgs framework.}

Between the standard option of an elementary Higgs boson, 
and the extreme case of Higgsless theories, there is the interesting 
class of models where a light scalar state emerges as a 
pseudo-Goldstone boson from a strongly-interacting framework
at a scale not far from $\Lambda_\chi$~\cite{Kaplan:1983fs}.

As we have seen in the previous section, a satisfactory description 
of EWPO without a light Higgs boson is possible, but it requires a 
non-trivial price to pay in terms of naturalness of the corresponding  
effective theory. This problem can be ameliorated if the 
effective theory includes a new light scalar state that, acting almost 
like an elementary Higgs, cuts off partially or completely 
the infrared-log  contributions to $S$ and $T$ in Eq.~(\ref{eq:STIR}).
Explicit examples of models of this type are the 
so-called Little Higgs models~\cite{Arkani-Hamed:2002qy} or 
the partially-composite Higgs models~\cite{Contino:2003ve}.
As pointed out in Ref.~\cite{Giudice:2007fh}, at low energies 
we can capture the main features of all this class of models
by means of an appropriate effective-theory, written in terms of 
the SM degrees of freedom and the new scalar field. 
The corresponding Lagrangian is nothing but the SM Lagrangian, 
with the explicit Higgs field, plus a series of higher-dimensional 
operators such as those shown in Eq.~(\ref{eq:QSQT}).

The two key parameters of this construction are $f$, 
the symmetry-breaking scale of the new dynamics above 
the electroweak scale ($\xi = v /f <1$), and the 
effective coupling $g_\rho$, which control the mass 
spectrum associated with the new dynamics: beside the light 
scalar, the lightest non-standard states are expected at the 
mass scale $m_\rho \sim g_\rho f$. These two parameters determine
the suppression of the non-standard higher-dimensional operators: 
in the simplifying limit where the new dynamics is 
maximally strongly coupled ($g_\rho \sim 4\pi$), 
the effective scale $\Lambda$ 
in Eq.~(\ref{eq:STloc}) can be identified with $\Lambda_f = 4\pi f$. 
For $\xi \to 0$ we recover the SM scenario (with no 
additional higher-dimensional operators) and, not 
surprisingly, we find an excellent fit of EWPO with
a light Higgs. However, it is clear that this is 
a fine-tuned limit. Even if we start with a vanishing Higgs 
mass at the tree-level, because the Higgs field is 
as Goldstone boson of the global symmetry breaking 
occurring at the scale $f$, radiative correction 
due to SM gauge interactions generate
an effective Higgs mass. This is typically of order 
\be
m^2_h \sim \frac{g^2  }{16\pi^2}  \frac{g_\rho^2  }{16\pi^2}  \Lambda_f^2
\sim \frac{g_\rho^2 }{16\pi^2} \frac{(180~{\rm GeV})^2}{\xi^2}~, 
\ee
hence $\xi^2$ can be considered as a good 
estimator of the fine-tuning of this class of models.
The situation can be improved if the new high-energy dynamics
is not maximally coupled ($g_\rho < 4\pi$).
However, in this case local contributions to EPWO maybe 
parametrically be enhanced by a factor of order 
$16\pi^2/g^2_\rho$ 
due to new light states (such as the vector resonances 
discussed in the previous section)~\cite{Giudice:2007fh}.
As a result, in all cases some amount of fine-tuning 
cannot be avoided.

On the phenomenological side, the most interesting 
aspect of this construction is the appearance of 
an effective Higgs boson, whose couplings receive 
corrections of $\cO(\xi^2)$ relative to their SM values.
While the fine-tuned case $\xi \ll 1$ will be very hard 
to distinguish from the SM, there are 
good experimental prospects to identify this 
scenario if $\xi =\cO(1)$~\cite{Grojean:2009fd}

\section{Composite fermions}

The last option we will briefly address is the possibility  
of adding new fermions, with well-defined transformation 
properties under $SU(2)_{L+R}$, as relatively light states 
below the na\"ive cut-off~$\Lambda_\chi$.
In the two frameworks discussed before we assumed that 
the SM fermions were coupled to the new dynamics in a 
weak and flavour-universal way, only via gauge interactions.
This {\em safe} assuption (as far as flavour physics 
is concerned) is questionable if we add new light 
fermion states which can mix with SM fermions.
On the one hand, this mixing opens the possibility of new 
potentially large custodial symmetry breaking terms 
(similarly to the top-quark mass), which could allow 
a good $S$--$T$ fit with no or heavy Higgs field.
On the other hand, this mixing leads to potential   
problems with other precision measurements, such as the 
$Z\to b \bar b$ coupling~\cite{Agashe:2006at}, 
the universality of charged-current interactions~\cite{Antonelli:2008jg,Bernard:2007cf}
and various flavour-changing neutral-current (FCNC) observables~\cite{Isidori:2009px}.
A natural way to deal with these problems is to enlarge
the symmetry group, possibly including some protective flavour 
symmetry, and to assume that the enlarged symmetry group 
is broken only along specific directions
(see e.g.~Ref.~\cite{D'Ambrosio:2002ex,Hirn:2004ze}).

As far as flavour-changing processes are concerned, 
the most protective assumption is the so-called MFV hypothesis. 
Within the quark sector, this implies 
that the global quark-flavour symmetry of the SM gauge Lagrangian,
${\mathcal G}_{q} = SU(3)_{Q_L} \times SU(3)_{u_R} \times SU(3)_{d_R}$, 
is broken only by two spurions (the Yukawa couplings)
transforming as 
$Y_U \sim (\bar 3, 3, 1)$ and $Y_D \sim (\bar 3, 1, 3)$
under~${\mathcal G}_{q}$~\cite{D'Ambrosio:2002ex}.
As recently discussed in~\cite{Barbieri:2008zt}, a MFV structure 
can be implemented if we assume that the new fermions 
are $SU(2)_{L+R}$ singlets. However, in such case 
the  contribtuion to $T$ is negligible. The impact in 
EWPO is potentially larger in the case of
$SU(2)_{L+R}$  dublets or triplets~\cite{Barbieri:2008zt}; 
however, in such case the maximal flavour symmetry 
group is smaller than ${\mathcal G}_{q}$
and we are left with fine-tuning problems 
in FCNC observables.

\section{Conclusions: the unavoidable fine-tuning problem}

As we illustrated with various examples, our knowledge about 
the ultraviolet completion of the effective Lagrangian in 
(\ref{eq:Luniv}), or the SM with no Higgs, is still quite poor. 
 
On the one hand, we have the class of models with a light Higgs 
field and relatively heavy new dynamics, whose extreme case is 
SM itself. In these models it is natural to obtain a good fit 
to electroweak precision observabels. However, there is an unavoidable 
fine-tuning problem in keeping the Higgs mass light. 
To some extent, this statement remains true even if the Higgs 
is a pseudo-Goldstone boson, or a partially composite state. 

On the other hand, we have the  class of models with 
no or heavy Higgs field. Here the fine-tuning problem on the Higgs mass 
is absent by construction; however, a naturalness problem 
re-appear in the conditions we need to impose on the free parameters 
of the effective theory in order to fit the 
electroweak precision observabels.

At present, comparing these two types of naturalness problems 
is more a question of taste than a well-defined physical question.
Hopefully, the problem will soon have a definite experimental answer 
with the direct exploration of the TeV energy range at the LHC.

\end{document}